# A natural NISQ model of quantum self-attention mechanism

Shangshang Shi [a], Zhimin Wang [a,*], Jiaxin Li [a], Yanan Li [a], Ruimin Shang [a], Haiyong Zheng [a], Guoqiang Zhong [a], and Yongjian Gu [a]

[a] Faculty of Information Science and Engineering, Ocean University of China, Qingdao, China

*Corresponding author: Zhimin Wang, wangzhimin@ouc.edu.cn

**Abstract**

The self-attention mechanism (SAM) has demonstrated remarkable success in various applications. However, training SAM on classical computers becomes computationally challenging as the number of trainable parameters grows. Quantum neural networks (QNNs) have been developed as a novel learning model that promises to provide speedup for pattern recognition using near-term Noisy Intermediate-Scale Quantum (NISQ) computers. In this work, we propose a completely natural way of implementing SAM in QNNs, resulting in the quantum self-attention mechanism (QSAM). The fundamental operations of SAM, such as calculating attention scores and producing attention features, are realized by only designing the data encoding and ansatz architecture appropriately. As these are the fundamental components of QNNs, our QSAM can be executed efficiently on near-term NISQ devices. Our QSAM models achieve better performance in terms of both accuracy and circuit complexity on the text categorization task. Moreover, the robustness of QSAM against various typical types of quantum noise is demonstrated, indicating the model's suitability for NISQ devices. The present QSAM will serve as the fundamental building blocks for developing large models of quantum attention neural networks for quantum advantageous applications.

## 1. Introduction

The self-attention mechanism (SAM) is a critical component of many powerful deep learning architectures that have achieved state-of-the-art performance across various tasks such as natural language processing [1-2] and computer vision [3-4]. Unlike convolutional and recurrent models, which process only one local neighborhood at a time, SAM is based on non-local operations [4], allowing it to capture long-range dependencies within the features and understand the data from a global perspective. This unique advantage has contributed to the excellent performance of SAM-based architectures [5]. However, SAM's quadratic complexity makes it computationally inefficient, severely limiting its applicability, especially when equipped with up to hundreds of billions of parameters, making it extremely challenging to train on classical computing resources [6].

In recent years, remarkable breakthroughs have been made in the field of quantum computing [7-9]. Notably, people are now entering the NISQ era of quantum computing, where NISQ refers to noisy intermediate-scale quantum computing [10]. There is a growing consensus that near-term NISQ devices will have advantageous applications by exploiting NISQ algorithms [11]. One of the most promising approaches is the development of quantum neural networks (QNNs) [12-13]. As a quantum analogue of classical neural networks, QNNs leverage parameterized quantum circuits (PQC) as a learning model [14], and can extend to deep neural networks with flexible multilayer architectures, such as quantum convolutional neural networks (QCNNs) [15-16] and quantum recurrent neural networks (QRNNs) [17-18].

QRNNs have the potential for more powerful learnability by utilizing the exponential large feature space generated by the PQC. To fully explore this huge space and capture long-range dependencies within the features, the SAM is necessary for QRNNs. However, currently there is no natural approach to implementing SAM in QNNs, where "natural" means that the mechanism should be integrated into the fundamental architecture of PQC. For reference, Cong et al. [15] proposed a natural NISQ model of QCNN that implements the convolutional and pooling mechanisms by appropriately designing the PQC. Developing a similar natural model of quantum self-attention mechanism (QSAM) is necessary for the advancement of QNNs.

So far, the following QSAM models have been proposed. Zhao et al. [19] introduced the quantum logical similarity (QLS) technique to evaluate attention scores and constructed a quantum self-attention neural network. While capable of efficiently computing attention scores, this model has a high circuit complexity, requiring several large auxiliary quantum registers and swap operations between registers. Li et al. [20] developed a quantum-classical hybrid self-attention neural network, where quantum circuits are mainly used to prepare *Query*, *Key*, and *Value* representations in the high-dimensional quantum feature space, while attention score calculations and weighted summation are implemented on the classical computer. A similar idea was also used by Sipio et al. [21] to construct a quantum Transformer. Such hybrid models only take quantum circuits as performance-enhanced versions of feedforward neural networks to replace those in the original classical models, and face interface problems between quantum and classical systems.

In this paper, we propose a natural and completely NISQ-compliant model of QSAM. Our main idea is based on the following fundamental observations: (1) the most basic operation in QNNs is implementing a PQC, which has adjustable quantum gates and can be trained to approximate almost any target function; (2) the key operations of SAM involve calculating attention scores and producing new features based on those scores. We find that the SAM operations can be naturally realized by properly designing the way of data encoding and the architecture of PQC. As a result, our QSAM is much more accessible on near-term NISQ devices.

The main contributions of this work are as follows:

(1) We develop a novel approach for implementing SAM in QNNs by combining the way of data encoding and PQC architecture. Our model is natural in terms of QNNs, because all the components required in this model are the properly-designed data encoding circuit and PQC which are most basic components of QNNs.

(2) The entire quantum circuit of our QSAM is hardware efficient, and can be customized concerning the gate set and connectivity of the quantum devices on hand. As a result, our QSAM can be executed efficiently on a variety of NISQ systems.

(3) Our QSAM achieves better performance compared to other quantum learning models on the task of text categorization using public datasets. In addition, our model is robust against various typical types of quantum noise, demonstrating its suitability for near-term NISQ devices.

The rest of the paper is organized as

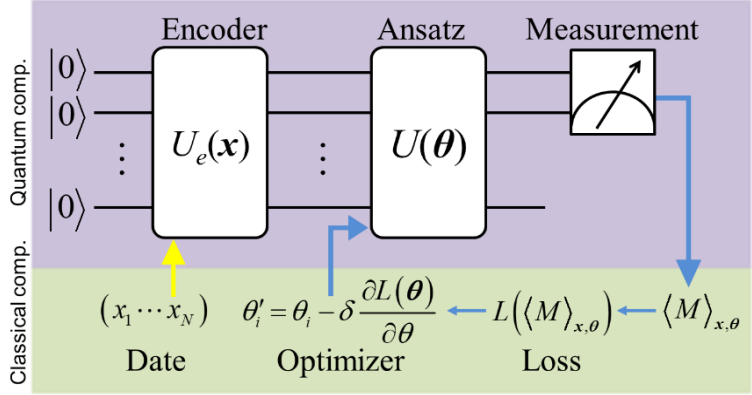

**Fig. 1.** The architecture of QNN. QNN is a hybrid machine learning model that combines quantum and classical computing, where the forward feature map is implemented on the quantum computer and parameter optimization is done on the classical computer.

follows. Section 2 reviews preliminary concepts about QNNs and SAM that will be used in this paper. Section 3 discusses the details of our QSAM architecture. Section 4 presents experimental results of QSAM's performance on tasks such as Iris flower classification and text categorization, as well as its robustness against quantum noise. Finally, we conclude our work in Section 5.

## 2. Preliminaries

### 2.1. Quantum neural networks

QNNs are quantum analogue of classical neural networks, which use PQC as machine learning models [22]. Generally, QNN consists of four parts: data encoder, ansatz, quantum measurement and parameter optimization routine, as shown in Fig. 1. The first three parts are implemented on the quantum computer, while the last optimization routine is carried out on the classical computer.

*The encoder circuit* is used to represent classical data as quantum states, which involves performing the unitary transformation $|x\rangle = U_e |0\rangle^{\otimes n}$, where $x$ is the data vector. Data encoding can be seen as a quantum feature map, which maps the data space to the Hilbert space where quantum states exist [23]. One of the commonly used encoding method is angle encoding that embeds data into the rotation angles of single-qubit or two-qubit rotation gates [24]. In the experiments of this paper, we use this method.

*Ansatz circuit* is a quantum analogue of classical feedforward neural network and is utilized to identify patterns within the feature generated by the encoder circuit. Ansatz is in fact a PQC consisting of adjustable quantum gates. Parameters in ansatz are optimized to approximate the target function that maps the feature to different value domains, each representing a distinct class.

There are generally two approaches to designing ansatz circuit, i.e. problem-inspired and hardware-efficient methods [22, 25-27]. The problem-inspired method requires the Hamiltonian related to the problem to construct the circuit, while the hardware-efficient method is directly assembling the native single-qubit and two-qubit gates based on the gate set and connectivity of the available quantum devices [27]. Therefore, hardware-efficient ansatz can be implemented efficiently on near-term NISQ devices.

*Quantum measurement* is utilized to generate a prediction or label for the inputting data [28]. The measurement corresponds to a physical observable $M$, that can be decomposed as $M = \sum_i \lambda_i |\lambda_i\rangle\langle\lambda_i|$, where $\lambda_i$ is the $i^{th}$ eigenvalue and $|\lambda_i\rangle$ is the corresponding eigenstate. After a measurement $M$, the quantum state $|\varphi\rangle$ will collapse to one of the eigenstates $|\lambda_i\rangle$ with the probability $p_i = |\langle\lambda_i|\varphi\rangle|^2$, and the expectation of measurement outcome can be expressed as

$$\langle M \rangle = \sum_i \lambda_i \cdot p_i = \sum_i \lambda_i |\langle\lambda_i|\varphi\rangle|^2. \quad (1)$$

The most frequently employed measurement in quantum algorithms is the computational basis measurement, namely the Pauli-Z measurement as follows:

$$\sigma_z = (+1)|0\rangle\langle 0| + (-1)|1\rangle\langle 1|. \quad (2)$$

***Optimization routine*** is used to update ansatz circuit's parameters, i.e. the rotation angles of gates. Much like classical models, the optimization process involves defining a loss function $L(\boldsymbol{\theta})$ and minimizing it in terms of the parameter vector $\boldsymbol{\theta}$. In QNN, commonly used loss functions such as mean squared error and cross-entropy loss can be utilized.

Similar to classical neural networks, parameters of ansatz can be updated using the gradient of the loss function. For instance, in the gradient descent method, we can update the $i^{th}$ parameter $\theta_i$ as follows:

$$\theta_i' = \theta_i - \delta \cdot \partial L(\boldsymbol{\theta})/\partial \theta_i, \quad (3)$$

where $\delta$ is known as the learning rate. It's worth noting that QNNs have no the backpropagation algorithm for evaluating gradient of parameters. Instead, the gradient is computed using techniques such as difference method or parameter shift rule on quantum computers [29-30].

### 2.2. Self-attention mechanism

SAM is a variant of attention mechanism that computes a representation of a sequence or image by relating different positions within it. In SAM, attention is calculated entirely based on feature vectors of the input data. Specifically, given a data sample $(\boldsymbol{x}_1,...\boldsymbol{x}_N)$ with $\boldsymbol{x}_i \in \mathbb{R}^d$, first three sets of feature vectors called *Query*, *Key* and *Value* are computed as

$$\boldsymbol{Q}_i = W_Q \cdot \boldsymbol{x}_i, \; \boldsymbol{K}_i = W_K \cdot \boldsymbol{x}_i, \; \boldsymbol{V}_i = W_V \cdot \boldsymbol{x}_i, \quad (4)$$

where $W_Q, W_K, W_V \in \mathbb{R}^{d \times d}$ represent the parameter matrices required to learn.

Then, the relevance of each *Key* position to the *Query* position is evaluated using a compatibility function. The resulting relevance score, namely attention score, represents the similarity or compatibility between the two positions, indicating how relevant a particular *Key* position is to the *Query* position. The two commonly used compatibility functions are additive function and dot-product function [31]. The dot-product function computes the attention scores as follows:

$$a_{ij} = \boldsymbol{Q}_i \cdot \boldsymbol{K}_j^T, \quad (5)$$

and the additive function uses the following way

$$a_{ij} = f(\boldsymbol{Q}_i, \boldsymbol{K}_j), \quad (6)$$

where $f(\cdot)$ is implemented using a feedforward neural network.

Finally, the attention at position $i$ is computed through a weighted sum of *Value* vector at all positions. Specifically,

$$\text{attent}_i = \sum_{j=1}^{N} \text{softmax}(a_{ij}) \cdot \boldsymbol{V}_j, \quad (7)$$

where the softmax function is used to normalize the $N$ attention scores. In practice, dot-product attention is faster and more space-efficient than additive attention because it can be implemented using highly optimized matrix multiplication code. The dot-product attention can be expressed in matrix form as [1],

$$\text{attent}(Q, K, V) = \text{softmax}\left(\frac{QK^T}{\sqrt{d}}\right)V, \quad (8)$$

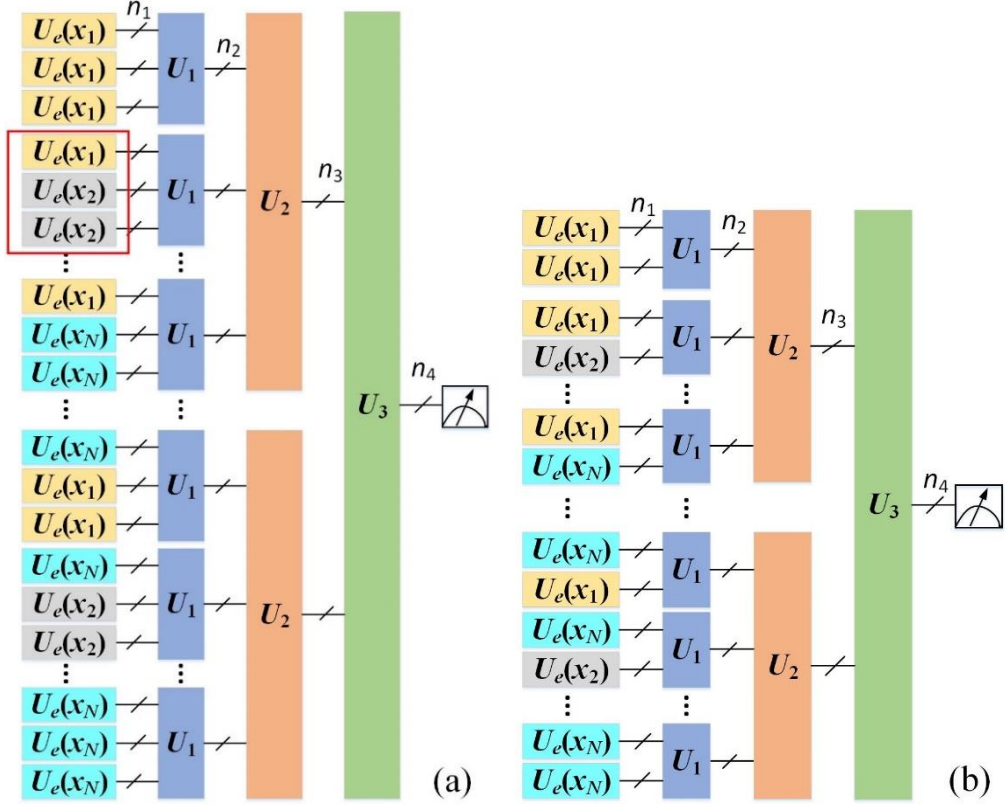

**Fig. 2.** Two architectures of QSAM, (a) the basic architecture (named QSAMb) and (b) the optimized version (named QSAMo). $U_e$ represents the encoding circuit, and $U_1$, $U_2$ and $U_3$ are the ansatz used to approximate target functions.

where $Q$, $K$ and $V$ are matrices that contain the *Query*, *Key* and *Value* vectors, respectively, and $\sqrt{d}$ is a scaling factor used to reduce the magnitude of dot products.

## 3. Quantum self-attention mechanism and associated networks

In quantum computing, the similarity between two quantum states can be characterized mathematically by their inner-product, which is typically evaluated using the SWAP test [32] or QLS method [33]. However, the resulting quantum circuits have high complexity to run on the near-term NISQ devices.

In this paper, we find that the attention vectors in the QSAM can be evaluated by properly design the way of data encoding and the architecture of PQC based on the basic spirit of classical SAM. Specifically, Eqs. (5), (6) and (7) can be generally formalized as

$$attent_i = \sum_{j=1}^{N} g(Q_i, K_j) \cdot V_j, \qquad (9)$$

where $g(\cdot)$ represents the dot product operation in Eq.(5) or feedforward transformation in Eq. (6). In QNN, we can put the vectors, $Q_i$, $K_j$ and $V_j$ together during the process of data encoding, and approximate the transformation $g(\cdot)$ and arithmetic operations through a PQC. Accordingly, the quantum circuit of our QSAM is designed as shown in Fig. 2, where Fig. 2(a) is the basic architecture denoted as QSAMb, and Fig. 2(b) is the optimized one denoted as QSAMo, which removes some redundant information from the QSAMb model. For example, in the encoding unit marked by the red box in Fig. 2(a), the data vector $x_2$ is embedded twice. In

$$\begin{bmatrix} att_1 \\ att_2 \\ \vdots \\ att_N \end{bmatrix} = \begin{bmatrix} Q_1 \\ Q_2 \\ \vdots \\ Q_N \end{bmatrix} \begin{bmatrix} K_1^T & K_2^T & \cdots & K_N^T \end{bmatrix} \begin{bmatrix} V_1 \\ V_2 \\ \vdots \\ V_N \end{bmatrix} = \begin{bmatrix} Q_1 K_1^T V_1 + Q_1 K_2^T V_2 + \cdots + Q_1 K_N^T V_N \\ Q_2 K_1^T V_1 + Q_2 K_2^T V_2 + \cdots + Q_2 K_N^T V_N \\ \vdots \\ Q_N K_1^T V_1 + Q_N K_2^T V_2 + \cdots + Q_N K_N^T V_N \end{bmatrix}. \quad (10)$$

order to reduce the circuit complexity of the model, one of the repeated encoding circuits could be left out, which can greatly decrease the number of qubits and trainable parameters. Below, we'll illustrate the model architecture in detail using the QSAMb.

As shown in Fig. 2(a), QSAMb begins with encoding circuits, i.e. $U_e(x_i)$. The encoding circuits serve two purposes. The first purpose is to obtain the *Query*, *Key* and *Value* vectors by embedding the classical data in the quantum feature space spanned by the encoding circuits. The second one is to put the $Q_i$, $K_j$ and $V_j$ vectors together in a manner as shown in Eq. (9). This is achieved using an encoding unit as shown by the red box in Fig. 2(a), where three encoding circuits constitute one encoding unit. It's worth noting that the quantum gates of encoding circuit can be fixed or adjustable (namely with or without trainable parameters) depending on the learning task.

The following $U_1$ and $U_2$ ansatz are used to implement the transformation $g(\cdot)$ and arithmetic operations in Eq. (9). When the $g(.)$ is the dot product operation, given a data sample $(x_1,\ldots x_N)$, Eq. (8) can be expressed as Eq.(10). Note that the first element of the attention vector $att_1$ equals to $Q_1 K_1^T V_1 + Q_1 K_2^T V_2 + \cdots + Q_1 K_N^T V_N$. Therefore, the $U_1$ circuit following the red box unit in Fig. 2(a) can be regarded as performing the operation $Q_1 K_2^T V_2$. The first $N$ $U_1$ circuits output the $N$ weighted value vectors, i.e. $Q_1 K_1^T V_1, Q_1 K_2^T V_2, \cdots, Q_1 K_N^T V_N$. Then the following $U_2$ circuit is used to approximate the operation $Q_1 K_1^T V_1 + Q_1 K_2^T V_2 + \cdots + Q_1 K_N^T V_N$ and obtain the first attention vector $att_1$.

When the $g(.)$ is the feedforward transformation, the functionality of $U_1$ and $U_2$ circuits can also be interpreted in terms of additive compatibility function. Specifically, the $U_1$ circuits correspond to the feedforward neural network to obtain the attention scores as shown in Eq. (6). By harnessing the strong correlation properties of the quantum circuits (both encoding and $U_1$ circuits), the feature vectors, not only the attention scores are obtained, i.e. $a_{ij}\tilde{V}_j = f_{U_1}(Q_i, K_j, \tilde{V}_j)$. Then the $U_2$ circuit is used to obtain the attention, i.e. $att_i = \sum_j a_{ij} \tilde{V}_j$.

The final layer of our QSAM model is made up of the $U_3$ circuit, which is used to transform the attention vectors. It is worth noting that the attention vectors can serve as input features for the $U_3$ circuit, and therefore, the design of the $U_3$ circuit can vary depending on the nature of the input feature. For instance, it can be designed using convolutional and recurrent architectures for image and sequential data, respectively. In the end, quantum measurement is applied to output a prediction for the classification or regression task.

Here we remark that all the data encoding circuits as well as the $U_1$, $U_2$ and $U_3$ circuits are implemented using the PQC. For the encoding circuits, PQCs with different structure would be adopted according to the properties of the learning data; whereas, $U_1$, $U_2$ and $U_3$ could be implemented using the PQCs with the same structure. The specific structure of the PQC is discussed in the section of numerical experiments.

## 4. Numerical experiments

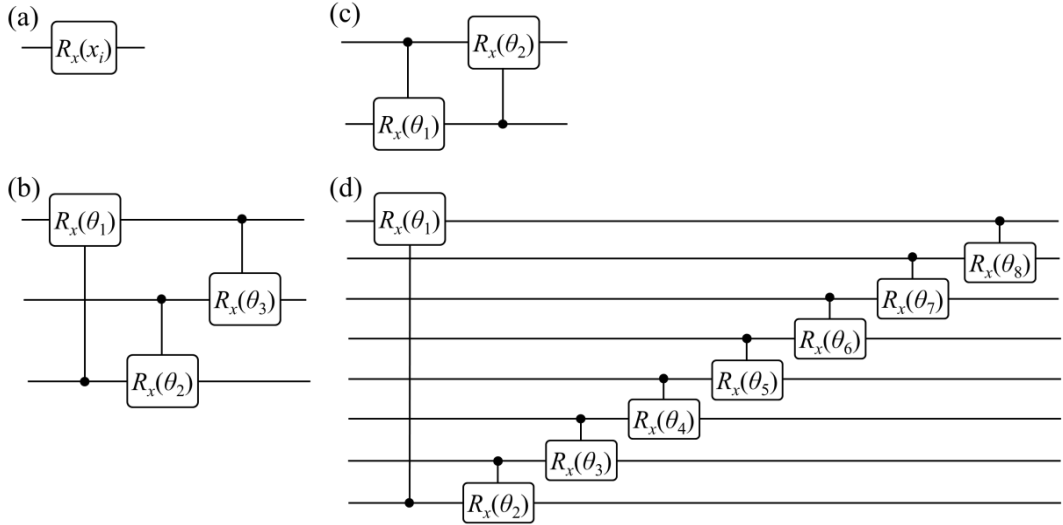

**Fig. 3.** The architecture of the quantum circuit for (a) data encoding unit $U_e(x_i)$, (b) $U_1$ transformation in QSAMb, (c) $U_1$ transformation in QSAMo, and (4) $U_2$ and $U_3$ transformation in both QSAMb and QSAMo. In these circuits, only the $R_x$ rotation gates are used.

**Table 1** Circuit complexity and the number of trainable parameters of QSAMb and QSAMo.

| Model | Complexity* | Trainable parameters |
|---|---|---|
| QSAMb | 48/88 | 19 |
| QSAMo | 32/72 | 18 |

\* The number of qubits and two-qubit quantum gates

The two QSAM models are verified concretely in the numerical experiments using three public datasets, namely the Iris flower data, as well as the MC and RP datasets for text categorization. The numerical experiments were conducted using Pennylane [34].

In the following experiments, we use the binary cross-entropy loss,

$$L(\boldsymbol{\theta}) = -\sum_{i=1}^{N}\left[\begin{array}{l}y_{true}\log\left(\Pr\left[f(\langle M\rangle_{x,\boldsymbol{\theta}})=1\right]\right)\\+(1-y_{true})\log\left(\Pr\left[f(\langle M\rangle_{x,\boldsymbol{\theta}})=0\right]\right)\end{array}\right] \quad (11)$$

where $y_{true} \in \{0,1\}$ is the class label and $\Pr[f(\langle M\rangle_{x,\boldsymbol{\theta}})=1]$ is the probability of the measurement outcome being 1. The Adam optimizer is utilized to update the parameters.

**4.1. Iris classification**

The first task involves the classification of Iris flowers into two categories [35]. The dataset consists of 100 data samples, with 50 samples per category for setosa and versicolour Iris. Each sample contains four attributes: sepal length, sepal width, petal length, and petal width. During the learning process, 80 samples were allocated to the training set, and the remaining 20 samples were used as the test set.

The encoding circuit $U_e(x_i)$ is implemented using only one $R_x$ rotation gate, as shown in Fig. 3(a). Therefore, according to the architectures shown in Fig. 2, the number of qubits $n_1$ equals to one. The $U_1$ circuits used in QSAMb and QSAMo are shown in Fig. 3(b) and 3(c), respectively, which utilize three and two qubits. In order to simplify the architecture of $U_2$ and $U_3$ circuits, two qubits of $U_1$ and $U_2$ circuits are utilized as the input to the following circuits (i.e. $n_2=2$, $n_3=2$). Since each Iris data sample contains four attributes, the number of qubits used by $U_2$ and $U_3$ circuits is eight. Fig. 3(d) presents the eight-qubits $U_2$ and $U_3$ circuits, which use the circuit-block configuration of two-qubit gates [36]. For binary classification problems, only one qubit needs to be measured (i.e. $n_4$=1). The circuit complexity of QSAMb and QSAMo as well as their number of trainable parameters are shown in Table 1.

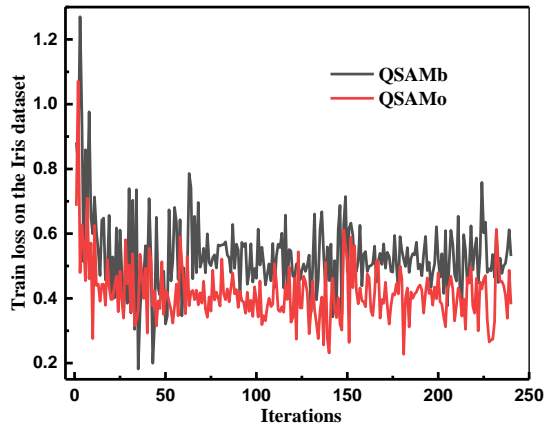

**Fig. 4.** Loss curves of the QSAMb and QSAMo models on the Iris dataset.

In the test using Iris dataset, both QSAMb and QSAMo models achieve a 100% classification accuracy, which indicates the two model's ability to uncover the internal relationships between the four attributes of Iris flowers and their categories. Additionally, Fig. 4 shows that QSAMo has a faster convergence rate than QSAMb and exhibits smaller loss values. These results suggest that reducing redundant input data effectively reduces the complexity of the QSAM model, enhancing its overall effectiveness.

### 4.2. Text categorization

The second task involves binary text categorization using two public datasets, i.e. the MC and RP synthetic dataset [37]. MC contains 130 sentences and each sentence has three or four words. Half of the sentences are related to food and half to information technology (IT). The task of MC is to categorize a sentence as food or IT. There are totally 17 different words in MC and part of the vocabulary is in common between the two classes, so the task is not trivial. The learning process utilizes 70 sentences for training, 30 for verification, and 30 for testing. On the other hand, RP contains 105 noun phrases containing relative clauses. There is an overall vocabulary of 115 words. The task of RP is also a binary-classification one, with the goal to predict whether a certain noun phrase contains a subject relative clause ("device that detects planets") or an object relative clause ("device that observatory has"). The learning process utilizes 74 sentences for training and 31 for testing. In addition, the size of vocabulary and the sparseness of words make this task more challenging than the MC task.

For language datasets, the word is firstly represented by the word vector, which is typically obtained in the process of pre-trained word embedding. In this paper, we utilize the encoding circuit $U_e$ ($x_i$) to embed words, and compute the corresponding *Query*, *Key*, and *Value* vectors. Fig. 5 shows the circuits of encoding unit for MC and RP datasets. The encoding units use two qubits, so the number $n_1$ in Fig. 2 equals to two and $U_1$ circuits in QSAMb (QSAMo) utilize six (four) qubits. As it is in the Iris dataset, two qubits of $U_1$ and $U_2$ circuits are utilized as the input to the following circuits (i.e. $n_2$=2, $n_3$=2). Since each sentence sample contains four words, the number of qubits used by $U_2$ and $U_3$ circuits is eight. The $U_1$, $U_2$ and $U_3$ circuits are constructed using a circuit-block configuration of two-qubit gates as shown in Fig. 3(d). Finally, one qubit is measured to predict a label (i.e. $n_4$=1).

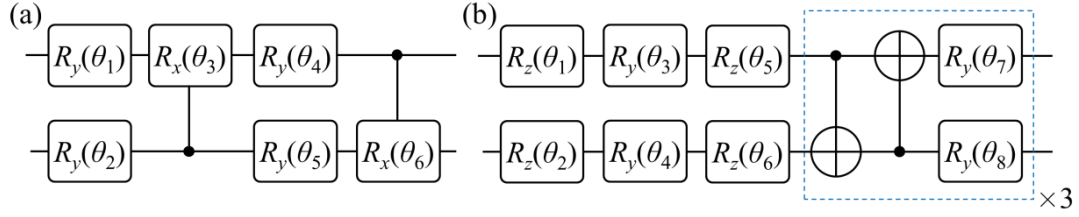

**Fig. 5.** Circuits of the encoder unit $U_e(x_i)$ for the dataset (a) MC and (b) RP. In MC, six parameters are used for word embedding, and in RP, 12 parameters are used.

**Table 2** Circuit complexity and the number of trainable parameters of four quantum learning models as well as their classification accuracy for the MC and RP dataset.

| Model | Complexity* | | Trainable parameters | | Accuracy (%) | |
|---|---|---|---|---|---|---|
| | MC | RP | MC | RP | MC | RP |
| DisCoCat[37] | — | — | 40 | 168 | 79.80 | 72.30 |
| QSANN[20] | 24/24 | 48/240 | 25 | 109 | 100.00 | 67.74 |
| QSAMb | 96/136 | 96/136 | 22 | 22 | 100.00 | 72.58 |
| QSAMo | 64/104 | 64/104 | 20 | 20 | 100.00 | 74.19 |

\* The number of qubits and two-qubit quantum gates

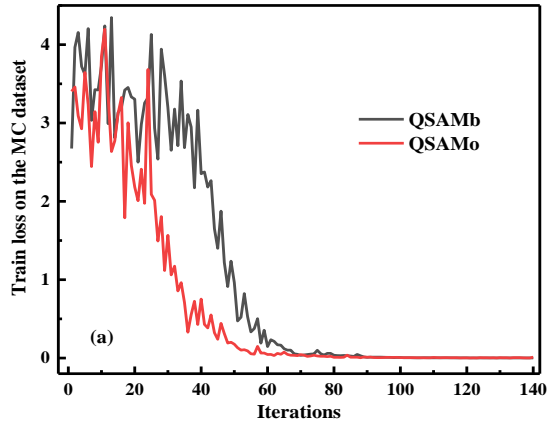

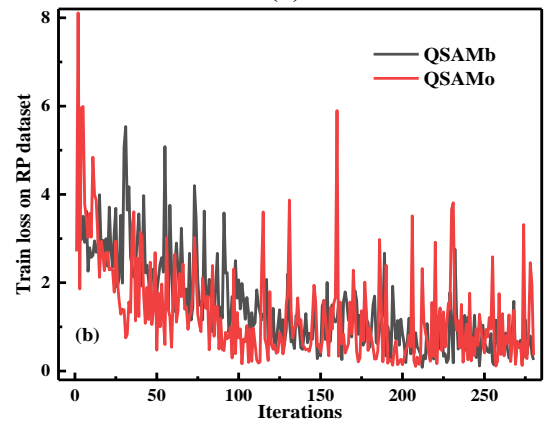

**Fig. 6.** Loss curves of the QSAMb and QSAMo models on the (a) MC and (b) RP dataset.

The circuit complexity of QSAMb and QSAMo as well as their number of trainable parameters are shown in Table 2. Note that the parameters in the encoder circuit are not counted in the total the number of trainable parameters as it is done in the QSANN mode [20], because the parameterized encoder could be replaced by pre-trained word embedding.

Table 2 shows that both the QSAMb and QSAMo models outperform other quantum learning models for NLP, such as DisCoCat [37] and QSANN [20]. In the MC task, both QSAMb and QSAMo achieve 100% classification accuracy; and in the RP task, QSAMb achieves 72.58% accuracy and QSAMo achieves 74.19% accuracy, both higher than the other two models. It's worth noting that the QSANN model is a quantum-classical hybrid model, where the PQC is used as a quantum-enhanced feedforward neural network to prepare *Query*, *Key*, and *Value* features, and other operations of SAM are implemented on the classical computer. Our QSAMs are complete quantum models. They achieve higher accuracy with significantly lower circuit complexity and fewer training parameters, demonstrating their powerful ability.

Additionally, Fig. 6 displays the training loss curves of the QSAMb and QSAMo models, which indicates that QSAMo has a faster convergence rate than QSAMb. This again suggest that reducing redundant input data can effectively enhance the model's effectiveness in text categorization task.

### 4.3. Robustness of QSAM

Finally, we demonstrate the robustness of our QSAM models against the quantum operation noises. For this experiment, we only use the QSAMo model due to its superior performance in terms of complexity, convergence and accuracy. Three typical types of quantum noise are used, namely the bit-flip, depolarizing noise and amplitude-damping, which can be respectively described by the following Kraus operators,

$$K_0^b = \sqrt{1-p}\begin{pmatrix} 1 & 0 \\ 0 & 1 \end{pmatrix}, K_1^b = \sqrt{p}\begin{pmatrix} 0 & 1 \\ 1 & 0 \end{pmatrix}, \quad (12)$$

$$K_0^d = \sqrt{1-p}\begin{pmatrix} 1 & 0 \\ 0 & 1 \end{pmatrix}, K_1^d = \sqrt{p/3}\begin{pmatrix} 0 & 1 \\ 1 & 0 \end{pmatrix},$$
$$K_2^d = \sqrt{p/3}\begin{pmatrix} 0 & -i \\ i & 0 \end{pmatrix}, K_3^d = \sqrt{p/3}\begin{pmatrix} 1 & 0 \\ 0 & -1 \end{pmatrix} \quad (13)$$

$$K_0^a = \begin{pmatrix} 1 & 0 \\ 0 & \sqrt{1-\gamma} \end{pmatrix}, K_1^b = \begin{pmatrix} 0 & \sqrt{\gamma} \\ 0 & 0 \end{pmatrix}. \quad (14)$$

Bit-flip is the most fundamental type of quantum noise, which has a certain probability $p$ of flipping the state of a qubit (using the operator $K_1^b$), and a probability of $1-p$ of leaving the state unchanged (using the operator $K_0^b$). Depolarizing noise is a more general form of quantum noise that compasses both bit-flip and phase-flip noise, providing a more comprehensive representation. Amplitude damping is used to model how interaction with the environment can impact the state populations of a qubit, with $\gamma \in [0,1]$ representing the probability of amplitude damping.

We perform noise experiments by applying bit-flip, depolarization, and amplitude damping noise to the $0^{th}$ qubit of $U_3$, which serves as the final measurement qubit. Table 3 shows the results, which reveal that the extent of noise's impact depends on the type of noise and the dataset being used. For a relatively simple learning task like Iris and MC, classification accuracy remains at 100% even when noise levels reach up to 0.1. However, for the RP task, a decrease in accuracy is observed as the noise level increases. Moreover, the bit-flip and amplitude damping noise have a greater impact than the depolarizing noise.

**Table 3** Test accuracy of QSAMo on the Iris, MC and RP datasets using noisy quantum circuits.

| Noise channel | Noise level $p$ | Accuracy (%) | | |
|---|---|---|---|---|
| | | Iris | MC | RP |
| Zero noise | 0 | 100 | 100 | 74.19 |
| Bit-flip | 0.001 | 100 | 100 | 74.19 |
| | 0.005 | 100 | 100 | 70.97 |
| | 0.01 | 100 | 100 | 58.06 |
| Depolarizing | 0.001 | 100 | 100 | 70.97 |
| | 0.005 | 100 | 100 | 64.51 |
| | 0.01 | 100 | 100 | 67.74 |
| Amplitude Damping | 0.001 | 100 | 100 | 64.52 |
| | 0.005 | 100 | 100 | 67.74 |
| | 0.01 | 100 | 100 | 61.29 |

These findings suggest that our QSAM model is robust against quantum noise, so long as it possesses sufficient expressibility to learn the dataset. In other words, for a specific task, the critical factor still lies in finding an appropriate architecture for the QSAM network to learn the dataset and achieve high accuracy while also possessing strong resistance against quantum noise.

## 5. Conclusions

In this work, we propose a natural NISQ way of implementing SAM in QNNs by properly designing the method of data encoding and the architecture of PQC. Specifically, two architectures are developed, namely the QSAMb and QSAMo model. The data elements corresponding to different positions are encoded in group to calculate the attention scores and the attention vectors are approximated using the ansatz circuit. QSAMo is obtained by removing redundant information from QSAMb. We evaluated the performance of the two QSAM models on several public datasets, including Iris classification and text categorization. QSAMo demonstrate a faster convergence rate than QSAMb and yield smaller loss values. In particular, our approach outperforms other quantum learning models in terms of accuracy and circuit complexity on the text categorization task. Moreover, we demonstrate the robustness of our model against quantum noise, such as the bit-flip, depolarizing, and amplitude-damping noise. The present work represents an important step of studying the quantum attention mechanism and the associated neural networks. The interesting future work includes (1) further optimizing the way of arranging the data encoding units to reduce the width complexity of the circuit; (2) optimizing the QSAM with better complexity and accuracy; (3) developing more complex quantum attention neural networks, such as the Transformer model, for various learning task based on the basic building blocks proposed in this work.

## Acknowledgments

We are grateful to the support of

computational resources from the Marine Big Data Center of Institute for Advanced Ocean Study of Ocean University of China.

**Funding**

The present work was supported by the Natural Science Foundation of Shandong Province of China [grant numbers ZR2021ZD19] and the National Natural Science Foundation of China [grant numbers 12005212].